\documentclass{article}

\usepackage{graphicx}
\usepackage{fleqn}
\usepackage{espcrc2}

\newcommand{\AmS}{{\protect\the\textfont2
  A\kern-.1667em\lower.5ex\hbox{M}\kern-.125emS}}

\title{
\vspace{-4.0cm} 
\begin{flushright}
{\normalsize\tt KEK-TH-645}\\
{\normalsize\tt RIKEN BNL Research Center preprint}\\
\end{flushright}
\vspace*{2.3cm}
SU(4) pure-gauge phase structure and string tensions\thanks{Talk
presented by SO.}\thanks{The authors thank RIKEN, Brookhaven
National Laboratory and the U.S.\ Department of Energy for providing
the facilities essential for the completion of this work.  SO
thanks the RIKEN BNL Research Center for its hospitality.
}}

\author{
Shigemi Ohta\address{Institute for Particle and Nuclear Studies, KEK,
Tsukuba, Ibaraki 305-0801, Japan}\({}^{\rm ,b}\)
and Matthew Wingate\address{RIKEN BNL
Research Center, BNL, Upton, NY 11973-5000, USA}
}

\begin{document}

\begin{abstract}
We present numerical evidence that the SU(4) pure-gauge dynamics has
a finite-temperature first-order phase transition.  For a \(6\times
20^3\) lattice, this transition occurs at the inverse-square coupling of
\(8/g^2 \sim 10.79\).  Below this and above the known bulk
phase transition at \(8/g^2 \sim 10.2\) is a confined phase in which we
find two different string tensions, one between the fundamental \({\bf
4}\) and \({\bf 4^*}\) representations and the other between the self-dual
diquark \({\bf 6}\) representations.  The ratio of these two is about
1.5.  The correlation in the adjoint representation suggests no string
forms between adjoint charges.
\end{abstract}

\maketitle

There are renewed interests in SU(\(N_c\)) pure Yang-Mills theory with
large \(N_c\):

1) Finite-temperature phase structure of quantum chromodynamics
(QCD) would be easier to understand if the SU(\(N_c\)) pure Yang-Mills
system has a second order phase transition for \(N_c \ge 4\)
\cite{PisarskiTytgat}.  With standard large-\(N_c\) analysis where
\(N_c g^2\) is held fixed, the Z(\(N_c\)) deconfinement transition occurs
at \(T_d \sim O(1)\), separating confining phase with free energy \(F \sim
O(1)\) and deconfining phase with \(F \sim O(N_c^2)\).
The deconfining temperature \(T_d \sim O(1)\) is not effected 
if \(N_f\) and \(g^2N_c\) are held fixed and \(N_c \rightarrow \infty\).
If the transition is first order, it is not effected either.
So large \(N_c\) is not a reasonable guide
for \(T\ne0\) QCD phase structure with Columbia phase diagram
\cite{ColumbiaPD}, unless SU(\(N_c\)) pure Yang-Mills dynamics has second
order deconfining phase transition for all \(N_c \ge 4\).

2) New developments in M/string theory \cite{Maldacena} predict
such things as glueball spectrum at large \(N_c\) and large \(g^2\) or
ratio between different string tensions  for \(N_c \ge 4\)
\cite{Strassler}.

3) The dimensionless ratio \(T_d/\sqrt{\sigma}\) of the
deconfining temperature \(T_d\) and string tension \(\sigma\) is expected
to be independent of \(N_c\) with a value\(\sqrt{3/\pi(D-2)}\) with \(D\)
being the space-time dimensions \cite{PisarskiAlvarez}.

Here we report the results of our numerical investigation on the order of
deconfining phase transition and the ratio of string tensions for
\(N_c=4\) \cite{Lattice98}.  We use single-plaquette action defined in the
fundamental \({\bf 4}\)-representation of the SU(4) gauge group. 
Combinations of pseudo-heatbath or Metropolis and over-relaxation
algorithms are used in updating 4, 6 and 8 \(\times 8^3\),
\(12^3\), \(16^3\) or \(20^3\) lattices.  Various workstations are
used for the numerical calculations, while migration to the RIKEN BNL
QCDSP mother boards is planned.  We look at the following observables:
plaquette, Polyakov loops, \(L(\vec{x}) = (1/N_c) {\rm tr}
\prod_{t=1}^{L_t} U(\vec{x},t; \hat{t})\), in \({\bf 4}\) (fundamental),
\({\bf 6}\) (antisymmetric diquark), \({\bf 10}\) (symmetric diquark) and
\({\bf 15}\) (adjoint) representations, deconfinement fraction, and
Polyakov loop correlation \(
\langle L(\vec{0}) L(\vec{r})^* \rangle
\sim r^{-1}\exp(-F(r)/T)
\sim \exp(-L_t\sigma r - \ln r)\) in \({\bf 4}\), \({\bf 6}\), \({\bf
10}\) and \({\bf 15}\) representations.

This SU(4) pure Yang-Mills system is known to have a bulk phase
transition near \(\beta = 8/g^2 \sim 10.2\), separating two confining
phases \cite{BCM}: across this transition the plaquette jumps
discontinuously but the average Polyakov line in the fundamental \({\bf
4}\) representation remains zero on both sides.  However if the lattice
extent in temperature direction \(L_t\) is too small this bulk transition
drives a first-order finite-temperature deconfining transition
just above itself in \(\beta\) \cite{GockschOkawa}.  We confirmed this is
indeed the case for \(L_t=4\).  But for \(L_t \ge 6\) we have good enough
separation between the bulk and finite-temperature transitions as desired.

As is shown in Figure \ref{fig:coex}, on a \(6 \times 20^3\) lattice we
confirmed coexistence of confined and deconfined phases at temperature
\(\beta\)=10.79:
\begin{figure}
\includegraphics[width=70mm]{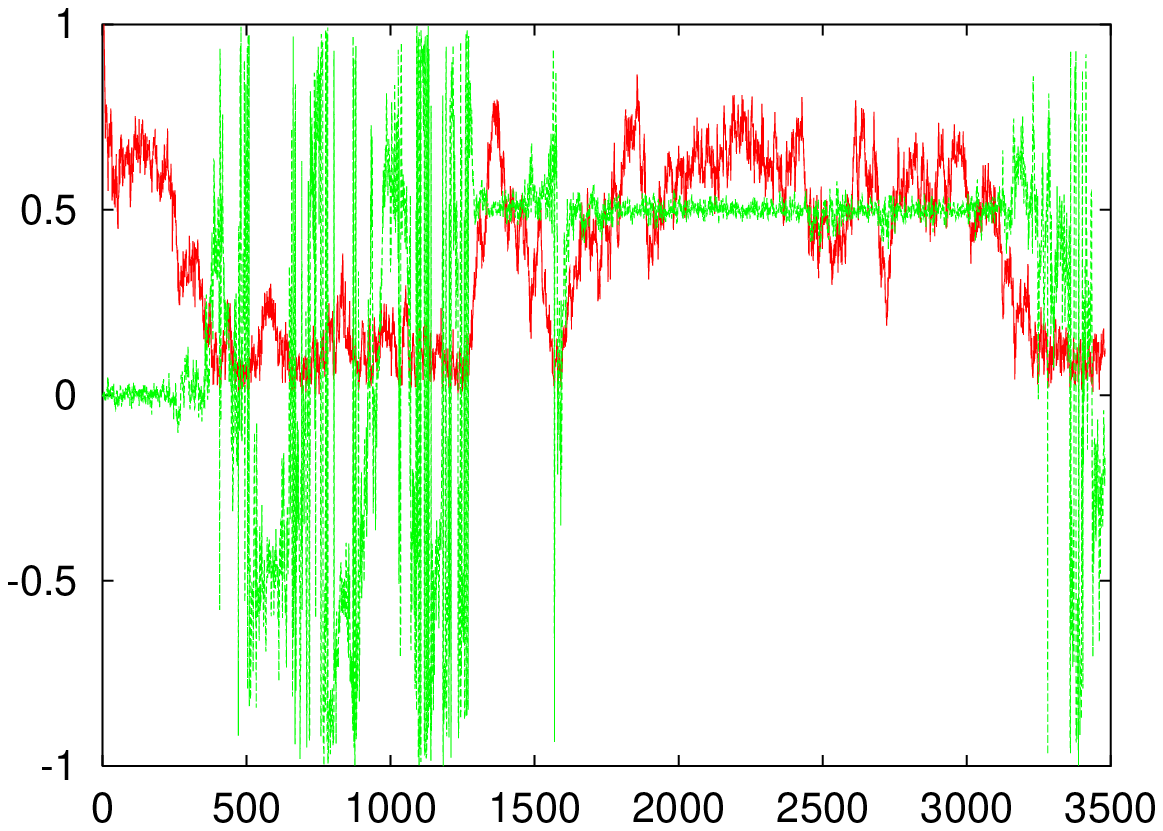}
\includegraphics[width=70mm]{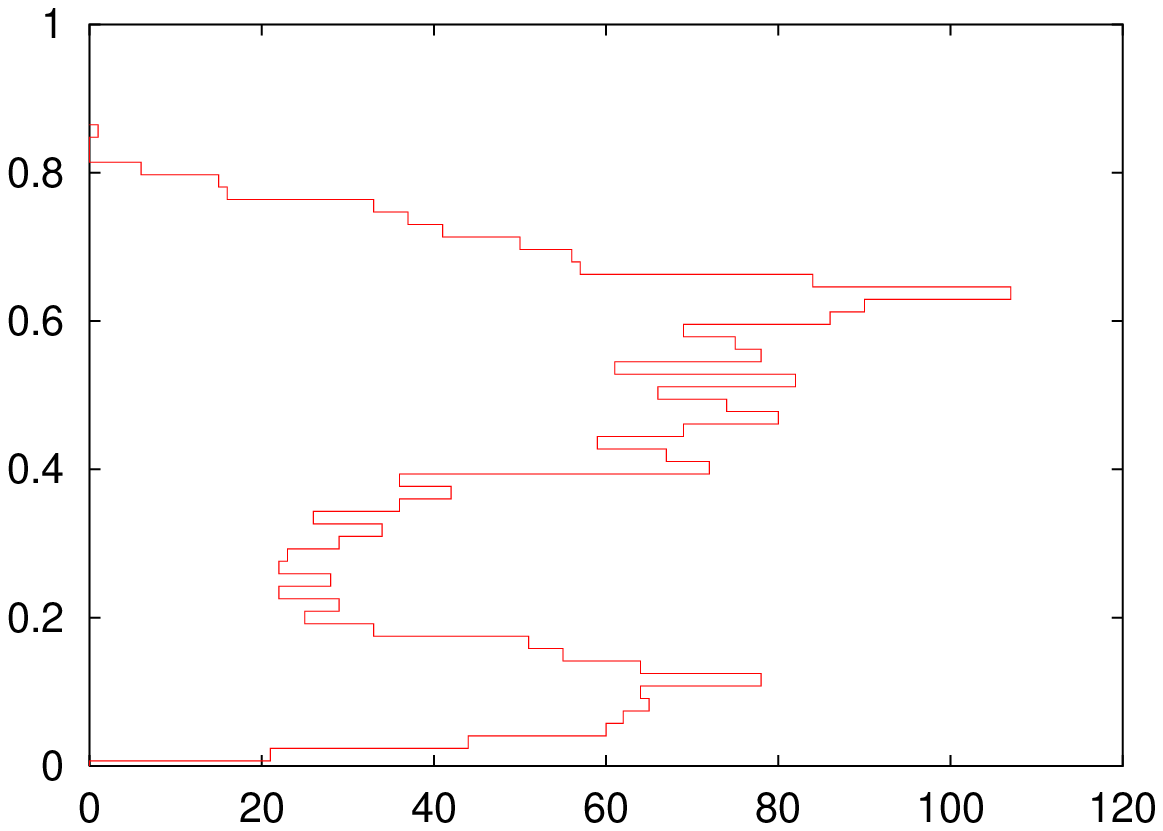}
\caption{Time histories of the fundamental Polyakov loop
magnitude and argument (in units of $\pi$) from a $6 \times 20^3$
lattice at $\beta=10.79$ (above) and magnitude histogram (below). 
Confined and deconfined phases coexist at this temperature suggesting a
first-order deconfining phase transition.}
\label{fig:coex}
\end{figure}
This strongly suggests a first-order deconfining phase
transition.  Work in progress confirms this phase coexistence
as we extend the simulation from the present 3500 evolution
steps (1 evolution = 5 heat bath + 1 over relaxation steps)
to 20000 steps.  We plan further study with finite-size scaling.

String tensions in SU(\(N_c\)) pure Yang-Mills system is classified by
its center Z(\(N_c\)) \(N_c\)-ality.  With \(N_c=4\), the fundamental
(\({\bf 4}\)) charge has 4-ality \(k=1\), the two diquark (\({\bf 6}\)
and \({\bf 10}\)) charges \(k=2\), and adjoint (\({\bf 15}\)) \(k=0\). 
The string tensions between these charges and their anticharges are
predicted to behave as
\(\sigma_k\) \(\propto\) \(\min\{k,N_c-k\}\)
by a standard strong-coupling analysis,
\(k(N_c-k)\)
by another strong coupling analysis \cite{Strassler},
and
\(\sin{(k\pi/N_c)}\)
by a SUSY strong coupling analysis \cite{Strassler}.  Generally the ratio
\(\sigma_k/\sigma_1\) should fall in the interval \(1 \le
\sigma_k/\sigma_1 \le 2\) \cite{Creutz}.  Note also that \(N_c=4\) is the
first example with different string tensions: in SU(3) pure Yang-Mills
system the fundamental (\({\bf 3}\)) and the symmetric diquark (\({\bf
6}\)) tensions are the same \cite{Ohta}.

\begin{figure}
\begin{center}
\includegraphics[width=65mm,bb=69 55 334 295]{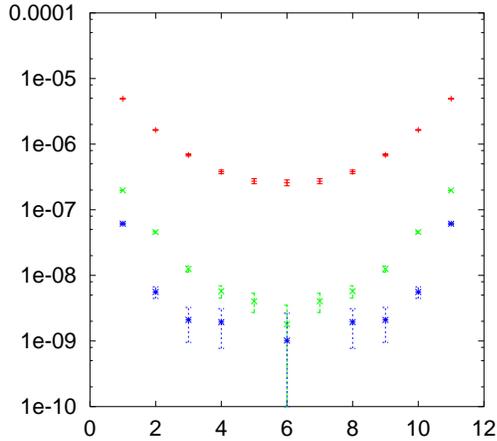}
\end{center}
\caption{Polyakov line correlation in ${\bf 4}$ ($+$), ${\bf 6}$
($\times$) and ${\bf 15}$ ($*$) representations on a $8\times
12^3$ lattice at $\beta$=10.85.  The adjoint (${\bf 15}$) signals now
suggest there is no string for this representation.  From the slopes of
the former two lower-dimensional representations we confirm different
string tensions for them, and by comparison there is no tension seen in
the adjoint representation.}
\label{fig:10.85}
\end{figure}

In our numerical calculation on a \(6\times 16^3\) lattice
at \(\beta=10.70\) (see Figure \ref{fig:10.70}):
\begin{figure}
\begin{center}
\includegraphics[width=65mm,bb=69 55 334 295]{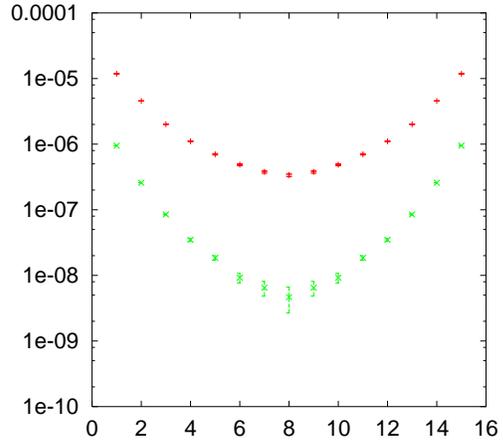}
\end{center}
\caption{Polyakov line correlation in ${\bf 4}$ ($+$) and ${\bf 6}$
($\times$) representations on a $6\times 16^3$ lattice
at $\beta=10.70$.  From the slopes we find different string
tensions for these representations.  No signal was obtained for
${\bf 10}$ and ${\bf 15}$ representations, probably because the
coupling is too strong.}
\label{fig:10.70}
\end{figure}
we find a clear difference between \({\bf 4}\)- and \({\bf 6}\)-string
tensions extracted from \({\bf 4}\)- and \({\bf 6}\)-Polyakov loop
correlations.  From fitting these data we have \(\sigma_4\)=0.068(4)
and \(\sigma_6\)=0.108(17).  At a stronger coupling of \(\beta\)=10.65
their values are 0.086(3) and 0.142(57) respectively.  Thus their ratio
\(\sigma_6/\sigma_4\) does not show much temperature dependence and falls
in the interval (1,2) as it should.  We are yet to see any signal for
\({\bf 10}\) and \({\bf 15}\) representations from this lattice, probably
because of too strong couplings.  On the other hand at a weaker coupling
of \(\beta\)=10.85 on a smaller lattice of \(8\times 12^3\) (see Figure
\ref{fig:10.85})
we find clear flattening of the adjoint (\({\bf 15}\)) correlation.

For thermodynamics of SU(4) pure-gauge theory we confirmed
that the bulk transition and \(T\ne0\) phase change are separated on
\(L_t\ge 6\) lattices, the bulk transition is at \(\beta_b \sim 10.2\)
and the finite-temperature 1st-order phase transition is at \(\beta_d
\sim 10.79\) (\(L_t=6\)) and \(\ge 10.9\) (\(L_t=8\)), and easier to
establish than weakly first-order SU(3).  For string tensions, at
\(L_t\)=6 we find signals for different string tensions in fundamental
\({\bf 4}\) and antisymmetric \({\bf 6}\) representations,  but no signal
yet for symmetric diquark \({\bf 10}\) and adjoint \({\bf 15}\)
representations.  These strings satisfy a relation,
\(
1 < \sigma_6/\sigma_4 <2,
\)
as they should, and the ratio does not show any strong
temperature dependence.  Combining these findings for
thermodynamics and string tensions at \(L_t\)=6, we find an inequality:
\(T_d/\sqrt{\sigma_4(T=0)}\) $<$ \(T_d/\sqrt{\sigma_4(T\sim T_d)}\) 
\(\sim\) 0.64 $<$ \(\sqrt{3/\pi(D-2)}\), just like in SU(2) and
SU(3) pure-gauge results.  At a weaker coupling of \(\beta=10.85\) using a
\(L_t\)=8 lattice we now have rather good evidence that there exists no
string in the adjoint (\({\bf 15}\)) representation.  We plan further
investigation on larger and finer lattices, probably using smaller
partitions of the QCDSP parallel supercomputer at RIKEN BNL Research
Center.

\end{document}